\documentclass[conference]{IEEEtran}
\IEEEoverridecommandlockouts
\usepackage{cite}
\usepackage{amsmath,amssymb,amsfonts}
\usepackage{algorithmic}
\usepackage{graphicx}
\usepackage{textcomp}
\usepackage{xcolor}
\usepackage{microtype}
\usepackage{booktabs}
\usepackage{paralist}
\usepackage{wasysym}
\usepackage{graphicx}
\usepackage{multirow,multicol}
\usepackage{amsthm}
\usepackage{mathrsfs}
\def\BibTeX{{\rm B\kern-.05em{\sc i\kern-.025em b}\kern-.08em
    T\kern-.1667em\lower.7ex\hbox{E}\kern-.125emX}}
\begin{document}

\title{CogIntAc: Modeling the Relationships between Intention, Emotion and Action in Interactive Process from Cognitive Perspective
}

\author{
	\IEEEauthorblockN{Wei Peng$^{1,2}$, Yue Hu$^{1,2*}$ \thanks{$*$ Corresponding author.}, Yuqiang Xie$^{1,2}$, Luxi Xing$^{1,2}$,  Yajing Sun$^{1,2}$}
	\IEEEauthorblockA{$^1$ Institute of Information Engineering, Chinese Academy of Sciences, Beijing, China}
	\IEEEauthorblockA{$^2$ School of Cyber Security, University of Chinese Academy of Sciences, Beijing, China}
	\IEEEauthorblockA{pengwei@iie.ac.cn, huyue@iie.ac.cn}
}


\maketitle

\begin{abstract}
Intention, emotion and action are important psychological factors in human activities, which play an important role in the interaction between individuals. How to model the interaction process between individuals by analyzing the relationship of their intentions, emotions, and actions at the cognitive level is challenging. In this paper, we propose a novel cognitive framework of individual interaction. The core of the framework is that individuals achieve interaction through external action driven by their inner intention. Based on this idea, the interactions between individuals can be constructed by establishing relationships between the intention, emotion and action. Furthermore, we conduct analysis on the interaction between individuals and give a reasonable explanation for the predicting results. To verify the effectiveness of the framework, we reconstruct a dataset and propose three tasks as well as the corresponding baseline models, including action abduction, emotion prediction and action generation. The novel framework shows an interesting perspective on mimicking the mental state of human beings in cognitive science.
\end{abstract}


\section{Introduction}
\label{intro}
In the process of human interaction, analyzing the relationships between IEA \{Intention, Emotion and Action\} is long-term research in Artificial Intelligence \cite{1997Affective,maslow2013theory}. Individual activities are the process of interacting with the external world driven by the intention \cite{1989Emergent,reeve2014understanding}. Specifically, the individual influences the external world via action \cite{minton2013belief}, while the external world influences the individual via emotion \cite{1989Emergent} to the individual. During this process, the individual's inner intention is the essence of the interaction \cite{bratman1987intention,Gable2010Approach}. The whole interactive process is realized and completed by the mutual cooperation of IEA. Analyzing and model the relationships between IEA at the cognitive level is of great significance to the in-depth understanding about the essence of human activities. 
It is also beneficial for various applications, including service satisfaction analysis \cite{montfort2000service}, intelligent agent \cite{padgham2005developing}, understanding intention \cite{DBLP:conf/aaai/SunST0DYSHS21}, emotion of consumers in live conversations \cite{DBLP:conf/ijcnlp/LiSSLCN17}, and so on. 

\begin{figure*}[t]
	\centering
	\includegraphics[width=0.95\textwidth]{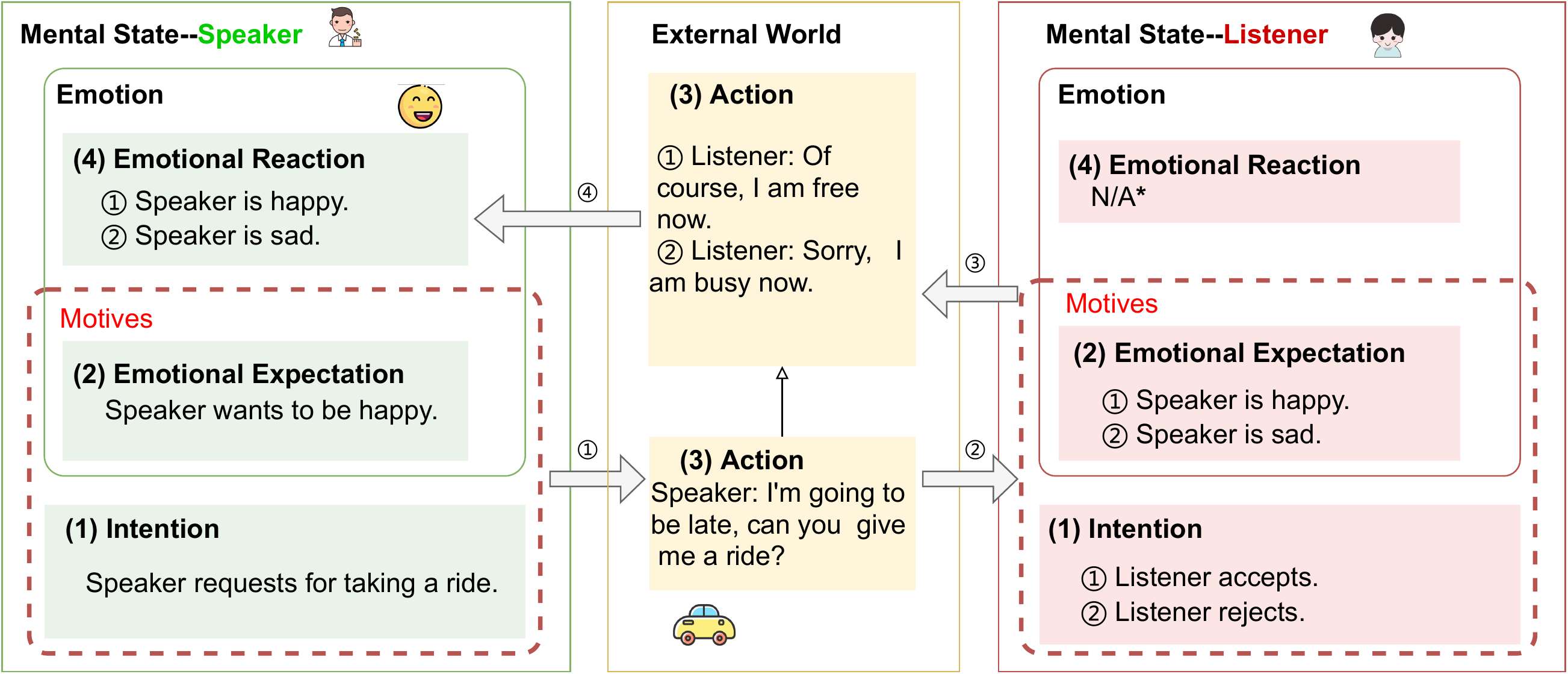}
	\caption{An example of interaction. N/A* means that the emotion has not yet been generated at this moment.}
	\label{fig:example}
\end{figure*}

Previous work \cite{DBLP:conf/emnlp/KumarAJ19,DBLP:conf/aaai/MajumderPHMGC19,DBLP:conf/aaai/ColomboCMVVC20,DBLP:conf/emnlp/GhosalMGMP20} mainly focus on emotion recognition or intention recognition of individuals, but seldom consider these factors of multiple individuals. Although some methods \cite {DBLP:conf/aaai/QinCLN020,DBLP:journals/corr/abs-2012-13260} propose to model the interaction between the emotion recognition task and intention recognition task, they just predict the label by the representation of utterances or graph attention network and doesn't take the inherent influence between IEA of multiple individuals into account. These works also lack analysis and explanation of emotional prediction and emotional cause. The RAIN \cite{9747565} consider the exploring the mutual relationships between intention and emotion, but ignoring the response generation during the interaction. 

To illustrate the process of interaction, an example is shown in Fig. \ref{fig:example} (the focus is on two individuals), where intention and emotional expectation constitute the motives that trigger the action, that is, driven by intention, the action is generated in the direction of emotional expectation. The definition of emotional expectation is the desired emotion towards achieving one's own goals. For the speaker, the goal is to accomplish his intentions, so the expectation is always positive. For the listener, the goal is to generate the action in the direction of speaker's emotional reaction, which in turn can be defined as the emotional expectation of the listener per se. Emotion can be further composed of emotional expectation and emotional reaction. When the speaker generates action, the listener reacts by generating corresponding motives, based on which, the speaker will generate the emotional reaction. The above process can be defined as \textit{interaction chain} {(e.g. \textcircled{1} $\rightarrow$ \textcircled{2} $\rightarrow$ \textcircled{3} $\rightarrow$ \textcircled{4})}. Before the speaker responds to the next turn at this moment, the listener will generate no emotional reaction. For the listener, the intention will be influenced by the speaker, and the intention \textit{accept or reject} will be produced. These two intentions correspond to two kinds of emotional expectations and actions, respectively. Namely, the listener's action is driven by the intention in the direction of speaker's emotional reaction. Finally, the speaker's emotional reaction is determined by whether the intention is satisfied. When satisfied, the speaker's emotion will be positive \cite{standage2005effect}, otherwise, non-positive.

We propose a novel Cognitive framework of individual InterAction (\textbf{CogIntAc}) to model the interaction between individuals. The framework contains three aspects: 
(1) The mental state \cite{DBLP:conf/acl/KnightCSRB18,DBLP:conf/acl/SmithCSRA18} between the speaker and listener is linked by actions to realize social interaction \cite{DBLP:journals/corr/abs-1904-09728}. (2) Intention is the essence of the action. The action is generated in the direction of emotional expectation under the drive of intention. By analyzing and understanding the action, the model can in turn obtain the intention via action abduction \cite{DBLP:conf/iclr/BhagavatulaBMSH20}. (3) Emotional reaction is determined by whether the action satisfies the intention, namely, it is a combined result of intention (mental state) and action (external world).

Around the CogIntAc, we have also proposed three novel tasks to discuss our framework. Considering that there is currently no dataset available, a dataset (see Sec.\ref{sec:data}) is manually created, which contains 2,106 single-turn dialogues. Each instance consists of utterances, intention and emotion labels and satisfaction labels.

The contributions can be summarized as follows:
\begin{compactitem}
	\item We propose a novel framework, CogIntAc, with the concept of \textit{a complete interaction} and \textit{interaction chain} in Sec. \ref{cogfr}, taking intention, emotion and action as the basic elements to model the process of human interaction. 
	\item Guided by the relationships between the nodes in the \textit{interaction chain}, three tasks and corresponding baselines are proposed, including action abduction of the speaker, emotion prediction of the speaker and action generation of the listener.
	\item We reconstruct a new dataset CogIEA (Intention, Emotion, Action) to verify the effectiveness of the framework.
	\item Experiments show that with the help of the CogIntAc, we can analyze the interaction in a deeper level and give a more reasonable explanation to the analysis results.
\end{compactitem}

\begin{figure*}[!]
	\centering
	\includegraphics[width=0.87\textwidth]{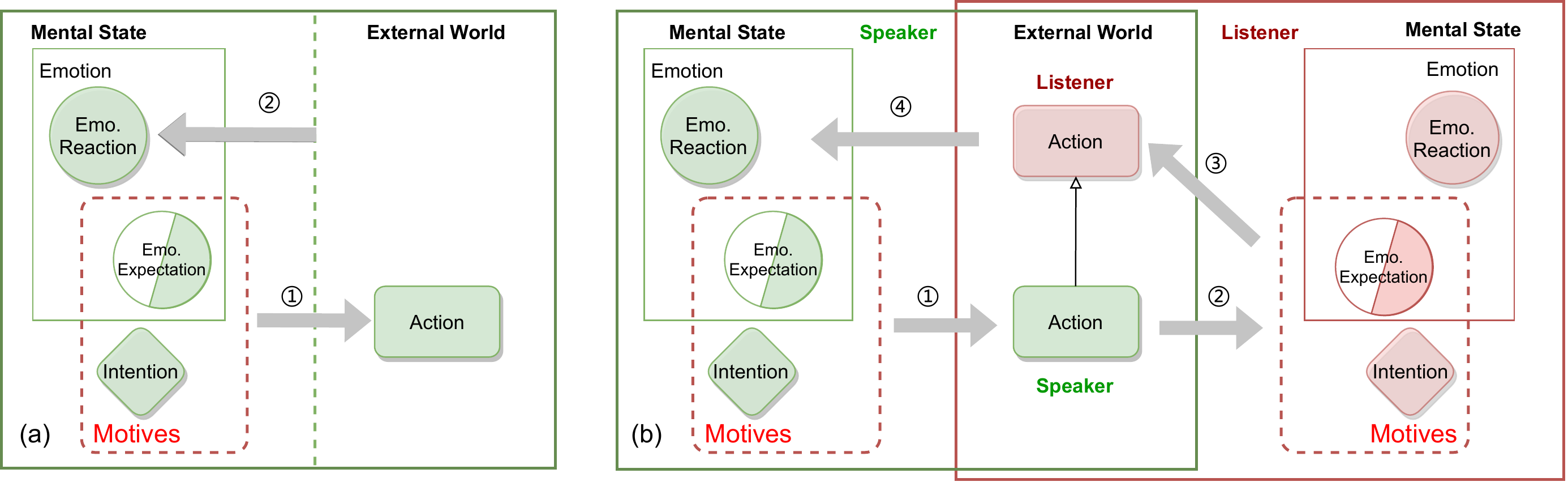}
	\caption{(a) Individual Activities Model (IAM). (b) An overview of proposed CogIntAc. Green indicates the speaker. The red one represents the listener. Diamond square indicates intention. The circle represents the emotional reaction and the semicircle represents the emotional expectation. The rectangle represents the action. The thick gray arrow represents the process. 
	}
	\label{fig:frame}
\end{figure*}

\section{Related Work}
\noindent
\textbf{Intention Recognition}
In dialogue systems, intention recognition requires models to predict the intention of the interlocutor.
With the development of deep learning,
\cite{DBLP:conf/emnlp/KumarAJ19} propose multi-task model for conversation modeling, which is optimized with dialogue act prediction and response selection.
\cite{DBLP:conf/aaai/ColomboCMVVC20} introduce a hierarchical encoder with attention mechanism for intention recognition. 
Few works \cite{2020Towards} consider emotion and only take the emotion classification as a joint task to boost intention recognition.
These methods primarily research individual intention recognition and ignore the influence of emotion on intention in mental state.

\noindent
\textbf{Emotion Recognition}
Previous studies \cite{DBLP:conf/emnlp/HazarikaPMCZ18,DBLP:conf/naacl/ColomboWMKK19,DBLP:conf/acl/WangZ18,DBLP:conf/aaai/Zhong0M19} show that in addition to the content of response, emotional communication between machine and human also plays an important role. 
\cite{DBLP:conf/naacl/HazarikaPZCMZ18} consider the role of inter-speaker dependency relations while classifying emotions.
\cite{DBLP:conf/aaai/MajumderPHMGC19} design a network to keep track of the individual party states throughout conversation and uses this information for emotion classification.
\cite{DBLP:conf/aaai/QinCLN020} propose DCR-Net to explicitly consider the cross-impact and model the interaction between the recognition of emotion and intention.
\cite{DBLP:conf/emnlp/GhosalMGMP20} incorporate different elements of commonsense for emotion recognition. 
These methods focus on recognizing an individual's emotion but also ignore the interaction between multiple individuals.

\noindent
\textbf{Response Generation}
There are plenty of works that study emotional response generation. 
The mainstream approach utilizes emotion-related factors, such as emotion vector, emotional memory, emotion dictionary, psychological intention, etc., to generate emotional responses.
For example,
\cite{asghar2018affective} propose three novel ways to incorporate emotional aspects into LSTM-based Seq2Seq neural conversation generation model.
Besides,
\cite{DBLP:conf/aaai/ZhouHZZL18} firstly propose to generate appropriate responses not only in content but also in emotion with three mechanisms. 
\cite{DBLP:conf/acl/SongZLXH19} generate meaningful responses with a coherent structure for a post, and express the desired emotion within a unified framework.
GLHG \cite{DBLP:journals/corr/abs-2204-12749} aims to capture the global cause, local intentions and dialog history, and model hierarchical relationships between them in the emotional support conversation \cite{DBLP:conf/acl/LiuZDSLYJH20}.
However, in psychology, the fundamental reason why human beings behave in the external world is that they take motivation as the starting point. 
Therefore, we focus on whether the generated response is consistent with our intention and in the direction of the emotion of speaker.

\section{Analysis of Individual Interaction}
The interaction between human beings is a process where humans interact through external world driven by their mental state. We first build the Individual Activity Model (IAM), and then introduce the interactive framework CogIntAc between individuals as well as the task definitions.
\subsection{Individual Activities Model}
\label{IAM}
Human individual activity is a process of interaction with the external world driven by mental state. In Fig. \ref{fig:frame} (a), the mental state includes intention, emotional expectation and reaction. The external world includes environmental factors and language actions, the main focus is on the language actions in this paper. \textcircled{\small{n}} indicates the information flows. The definitions of these elements are as follows:

\noindent
\textbf{Intention} As the psychological background to stimulate and guide action, intention is the essence of activity. In this paper, it particularly refers to the interaction intentions \cite{DBLP:conf/ijcnlp/LiSSLCN17}.

\noindent
\textbf{Emotional Expectation} The definition of emotional expectation is the desired emotion towards achieving one's goals, as demonstrated in Sec. \ref{intro}.

\noindent
\textbf{Emotional Reaction} It is the psychological reaction of an individual's inner satisfaction with the external world, which is formed by intention (internal causes) and external world (external causes). 

\noindent
\textbf{Action} An individual's words and deeds that influence the external world driven by his own intention.

According to the correlation of various elements above, the following relationships exist. (1) Intention v.s. Action: action can be predicted with the intention and emotional expectation. Intention can be analyzed by action abduction. (2) Emotional Reaction v.s. Intention: Emotional reaction can be predicted from the intention and external world.

\subsection{CogIntAc}
\label{cogfr}
From the perspective of interaction, we analyze the relationships between intention, emotion and action, and propose the CogIntAc based on IAM in Sec. \ref{IAM}. We define the concept of \textit{a complete interaction} and \textit{interaction chain}. The framework will be introduced in detail from four aspects.

\noindent
\textbf{CogIntAc Composition}
speaker and listener interact with each other in CogIntAc. As shown in Fig. \ref{fig:frame} (b), each of them is represented as an IAM structure, as introduced in \ref{IAM}, including four basic elements. Specifically, the intention classification has expanded into seven categories based on \cite{DBLP:conf/ijcnlp/LiSSLCN17}, including \textit{request, suggest, command, accept, reject, question, inform}. Emotional expectation is usually marked as happiness or content for speaker, as for listener, it is marked as the emotional reaction of the speaker. We have six labels of emotional reaction in CogIEA (e.g., \textit{happy, content, neutral, sadness, anger, disgust}).

\noindent
\textbf{Individual Interaction Process} 
The interaction process is defined as \textit{a complete interaction}, starting with the speaker's intention and ending with the speaker's emotional reaction. Specifically, starting from the intention, the speaker generates action in the direction of his emotional expectation driven by the intention. Then, the listener generates the corresponding motives to respond. Finally, the speaker generates an emotional reaction based on that response.

At the same time, we propose a new concept \textit{interaction chain} \textbf{(e.g. motives$_s$ $\rightarrow$ action$_s$ $\rightarrow$ action $_r$ $\rightarrow$ emotional reaction$_r$)} (subscript $s$ and $r$ indicate speaker and listener), which is a logical sequential chain between speaker and listener in CogIntAc, as shown by the gray arrows in Fig. \ref{fig:frame} (b). \textcircled{\small{n}} indicates the information flows. Each node in the \textit{interaction chain} has an impact on the final emotional reaction. 

\begin{figure*}[!]
	\centering
	\includegraphics[width=0.99\textwidth]{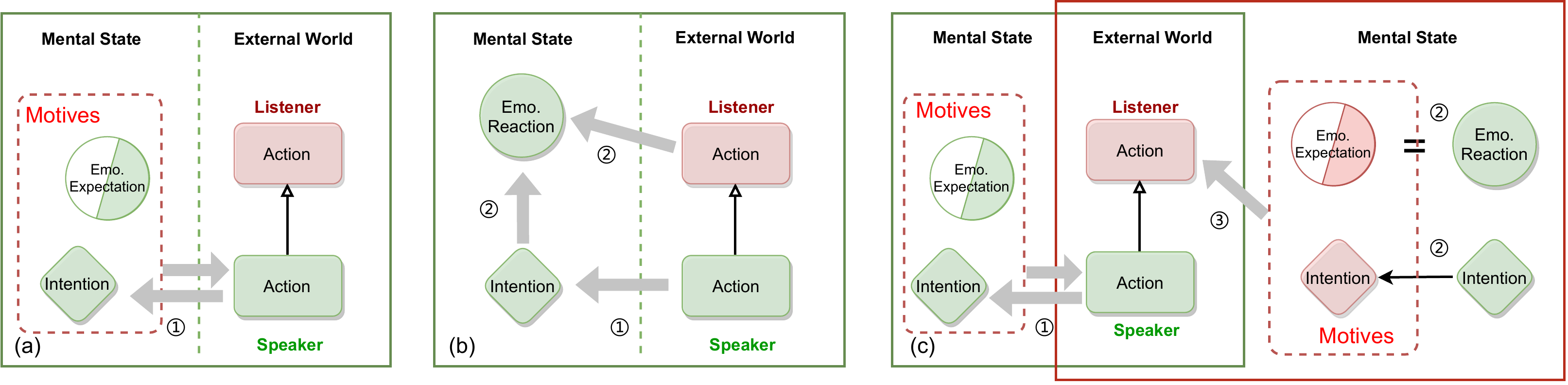}
	\caption{A diagram of the relationships between mental state and external world.
		(a) Action abduction of the speaker, the intention can be obtained by action. (b) Emotion prediction of the speaker. (c) Action generation. Intention and emotional expectation lead to the action of the listener.
	}
	\label{fig:relation}
\end{figure*}

\subsection{Relationships Analysis}
\label{relation_ana}
Since the interaction between individuals develops in line with the \textit{interaction chain}, we can reveal the essence of how two individuals complete the interaction, and give a reasonable explanation for the analysis results by analyzing the causes of each node on the \textit{interaction chain}. According to different roles of speaker and listener in the interaction process, we focus on analyzing the relationships between the speaker's intention, emotion and action, and predicting the listener's action. The subscript $s$ and $r$ indicate speaker and listener.

\noindent
\textbf{Action v.s. Intention of the speaker}
As shown in Fig. \ref{fig:relation} (a), the gray arrows indicate following relationships. (1) Action can be predicted when intention and emotional expectation are known. (2) intention can be obtained by action abduction. In this paper, considering that the action of speaker is often observable, the main focus is on the latter in predicting the intention of speaker.

\noindent
\textbf{Intention v.s. Emotion of the speaker} 
As shown in Fig. \ref{fig:relation} (b), the gray arrows indicate the relationship. The emotion$_s$ can be inferred when the intention$_s$ and the action$_r$ are known.

\noindent
\textbf{Action v.s. Motives of the Listener}
As shown in Fig. \ref{fig:relation} (c), the gray arrows indicate the relationship. It consists of two parts, motives inference of the listener and action generation of the listener. The intention$_r$ is influenced by intention$_s$, the emotional expectation$_r$ can be represented with the emotional reaction$_s$. And then, the action$_r$ is generated in the direction of emotional expectation$_r$ under the drive of intention$_r$.

The action$_r$ in the \textit{interaction chain} determines the final emotional reaction$_s$. Meanwhile, the analysis and prediction of the action$_r$ are of great significance in the decision-making of response strategy. 

\subsection{Task Definition}
\label{task_def}
Three tasks are proposed for relationships analysis in Sec. \ref{relation_ana}, including Action Abduction, Emotion Prediction and Action Generation.

\noindent
\textbf{Action Abduction}
Task Description: Given the action$_s$, the intention$_s$ is predictable.

\noindent
Task Input/Output: Action$_s$ is as input, intention of the speaker is as output.

\noindent
\textbf{Emotion Prediction} 
Task Description: Given the action$_s$ and the action$_r$, the task aims to predict the emotional reaction of the speaker.

\noindent
Task  Input/Output: Action$_s$ and action$_r$ are as input, emotional reaction of the speaker is as output.

\noindent
\textbf{Action Generation} 
Task Description: Given the action$_s$, emotional reaction$_s$, the task aims to predict the action of the listener.

\noindent
Task  Input/Output: Action$_s$, emotional reaction$_s$ are as input, action of the listener is as output.

\begin{figure*}[!]
	\centering
	\includegraphics[width=1.0\textwidth]{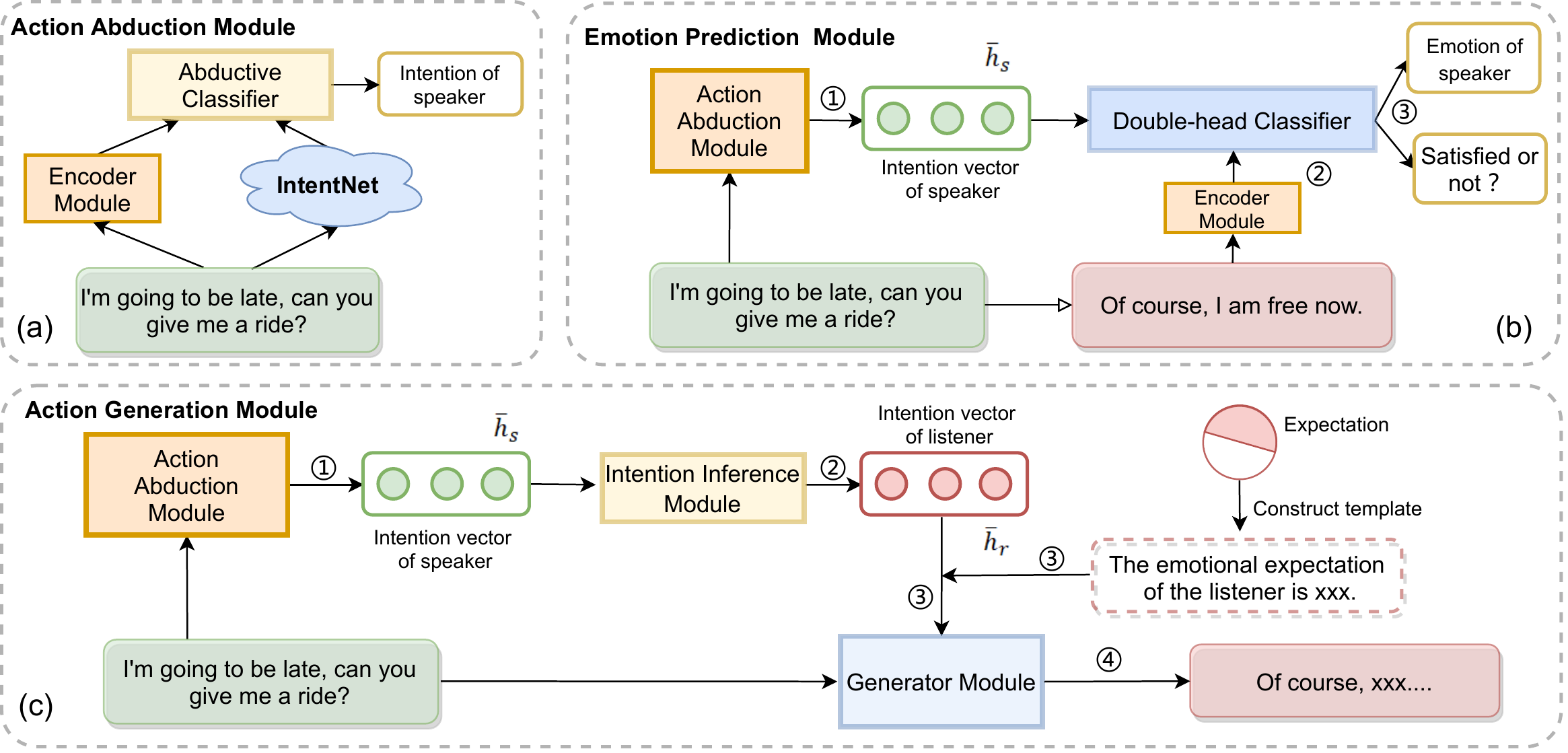}
	\caption{An overview of the proposed approach. (a) Action Abduction Module of the speaker. (b) Emotion Prediction Module of the speaker. (c) Action Generation Module of the listener. 
	}
	\label{fig:model}
\end{figure*}

\section{Model}
\label{sec:Model}
We design three components for each of the three tasks in Sec. \ref{task_def}. The model architectures for the tasks
are shown in Fig. \ref{fig:model}. Before proceeding, the formulation used in this paper is first described. Given a conversation $C = (u^s, u^r)$ a set of two utterances (actions) and $u^i = ({x^i_1, x^i_2, \dots, x^i_T})$ that consists of a sequence of T words,  $i \in \{s, r\}$, with $Y = (y^s, y^r)$ being the corresponding labels of intention and ${y_e^s}$ being the corresponding emotional reaction label of the speaker and ${\hat{y}}$ being the corresponding satisfaction label, and $s \in \{speaker\}$, $r \in \{listener\}$.
\subsection{Action Abduction Module}
\label{aam}
In this section, an Action Abduction Module is proposed to obtain the intention of the speaker. For the given utterance $u_1$, an encoder (LSTM model or pre-trained language models) models the contextual semantic information, using the first hidden state $\boldsymbol{h_s}$ as the representation of the intention of the speaker, and then with a \textit{Multilayer Perceptron (MLP)} to output probability distributions, as:

\begin{align}\label{equ:1}
\beta = {\rm Softmax}({\rm MLP}(\boldsymbol{h_s}))
\end{align}

\noindent
\textbf{Intention Dictionary} 
Observing that certain specific words such as \textit{ask for, proposal} can reflect intention, we extract keywords and construct an Intention Dictionary (IntDic) as the intention knowledge base by statistical method.
The IntDic outputs a probability distribution $\alpha$ of the keyword over all the corresponding intentions.

\noindent
\textbf{Abductive Classifier} 
Finally, with the Abductive Classifier (can be regarded as \textit{MLP}), the probabilities $\beta$ and $\alpha$ are jointly considered to obtain the speaker's intention by \textit{MLP}. We optimize the average cross entropy loss between all examples.

\subsection{Emotion Prediction Module}
\label{model:emo}
For emotion prediction task, the model predicts the emotional reaction of the speaker. As shown in Fig. \ref{fig:model} (b), we first model the intention of the speaker to obtain the intention vector $\boldsymbol{\bar{h_s}}$. This part has been implemented in Sec. \ref{aam}, as:
\begin{align}\label{equ:2}
\boldsymbol{\bar{h_s}} = {\rm ReLU}(W_{e}^\top [\boldsymbol{h_s}; \alpha])
\end{align}
where $W_{e}\in\mathbb{R}^{ (h+v) \times h}$, $[;]$ indicates vector concatenation. $v$ represents the dimension of the $\alpha$. 

Similarly, the hidden state of the action of the listener $\boldsymbol{h_r}$ is obtained with the encoder module.

\noindent
\textbf{Double-head Classifier} 
A Double-head Classifier is designed for predicting the emotion of the speaker and determining whether the intention of the speaker is satisfied. A fusion mechanism \cite{DBLP:conf/acl/WangWY18,DBLP:journals/kbs/PengHYXX21} is introduced for better fusing the intention of the speaker and action of the listener:
\begin{align}\label{equ:3}
\boldsymbol{f} = {\rm {Fuse}}(\boldsymbol{h_r}, \boldsymbol{\bar{h_s}})
\end{align}

Then, we consider two \textit{MLPs} for the multi-task learning, the average cross entropy loss is also optimized by Adam \cite{DBLP:journals/corr/KingmaB14}. Furthermore, the interpretation template is constructed to explain the cause of emotion, namely, utilizing the template to generate a more reasonable explanation. For example, the speaker's emotion is \textit{happy} because his intention is \textit{satisfied} by the listener.

\subsection{Action Generation Module}
For the listener, the model in Fig. \ref{fig:model} (c) predicts the action of the listener. Under the guidance of CogIntAc, we first make an inference to the intention of the listener by \textbf{Intention Inference Module}, because the intention of listener leads to the action. As for the emotional expectation, it is defined as emotional reaction of the speaker. Similarly, we construct the template \textit{The emotional expectation of the listener is xxx} to enrich the complete semantic information. Finally, the \textbf{Generator Module} (BART, GPT-2) outputs a response.

\noindent
\textbf{Intention Inference Module} 
The module uses a MLP to transform the intention space from the speaker to the listener, as:
\begin{align}\label{equ:4}
\boldsymbol{\bar{h_r}} = ({\rm MLP}(\boldsymbol{\bar{h_s}}))
\end{align}

\begin{table*}[!]	
	\centering
	\resizebox{0.92\linewidth}{!}{
		\begin{tabular}{lccccccccc}
			\midrule
			\multicolumn{1}{c}{\multirow{2}{*}} & \multicolumn{3}{c}{{Action Abduction}}       & \multicolumn{3}{c}{{Emotion Prediction}}        & \multicolumn{3}{c}{{Satisfaction Prediction}}  \\  
			\multicolumn{1}{c}{} 	& {P}	& {R}	& {F1}  	& {P} & {R}     & {F1}    & {P}	& {R}	& {F1}        \\ \hline
			{GRU} \cite{DBLP:journals/corr/ChungGCB14}	 	& {46.32}	& {41.27}  & {43.65} 	& {42.47} & {41.13}  & {41.79}    & {72.20} & {72.59}  & {72.39}    \\
			{GRU+Attention} \cite{Vaswani2017AttentionIA}		& 48.17	& 41.53	& 44.60 	& 43.55   & 41.85   & 42.68   & {73.16} & {74.55}  & {73.84}    \\
			{BERT} \cite{DBLP:conf/naacl/DevlinCLT19}		& 65.28	& 65.73	& 65.50 	& 55.14		& 55.49	& 55.31	 & {81.77} & {81.56}  & {81.66} \\ 
			{RoBERTa$_{base}$} \cite{DBLP:journals/corr/abs-1907-11692}  & 68.00   & 67.71  	& 67.85 	& 57.05 	& 57.48	& 57.26	 & {83.05} & {82.87}  & {82.95} \\	\hline
			{RoBERTa$_{large}$} \cite{DBLP:journals/corr/abs-1907-11692}		& {71.88}	& {70.63}	& {71.25}	& {58.68} & {59.05} & {58.85} 	 & {85.94} & {85.92}  & {85.93} \\
			\quad \textbf{+IntDic}		& {73.25}	& {72.19}	& {72.71}	& {60.81} & 61.53 & {61.16} 	 & {88.10} & {88.18}  & {88.14} \\
			\quad\textbf{+Fusion Mechanism} & {-}	& {-}	& {-}	& {58.37} & {61.41} & {59.85} 	 & {88.84} & {88.64}  & {88.74} \\
			\quad\textbf{+Multi-task} 	& {-}	& {-}	& {-}	& {59.79} & {59.84} & {59.81} 	 & {87.88} & {88.33}  & 88.11{} \\
			\quad\textbf{+All} 	& \textbf{73.25}	& \textbf{72.19}	& \textbf{72.71}	& \textbf{64.80} & \textbf{62.21} & \textbf{63.47} 	 & {\textbf{89.69}} & {\textbf{89.76}}  & {\textbf{89.72}} \\	\hline
			\textbf{Human Performance}	& 91.42	& 89.58	& 90.49	& 87.46 	& 86.15	& 86.80	 & {93.65} & {93.57}  & {93.60}  \\ \midrule
	\end{tabular}}
	\caption{\label{tab:main} Experimental results on CogIEA test set for tasks of Action Abduction and Emotion Prediction of the speaker. Human performance results are obtained by three testers. P indicates Precision, R means Recall. }
\end{table*}

\begin{figure*}[!]
	\centering
	\includegraphics[width=0.9\textwidth]{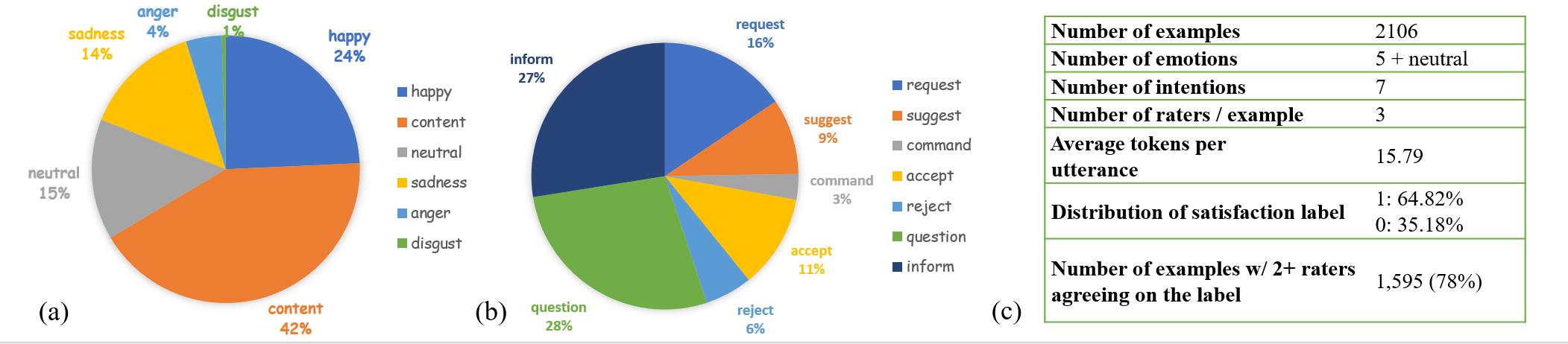}
	\caption{Distribution of emotion (a) and intention (b) in CogIEA. (c) The detail statistics of the dataset CogIEA. }
	\label{fig:data}
\end{figure*}

\noindent
\textbf{Generator Module} 
The generator module generates response by using a combination of intention $\boldsymbol{\bar{h_r}}$, emotional expectation template sentence $m_j$ and action of the speaker $x^s_i$, as:
\begin{equation}\label{equ:decoder}
p(y_t | y_1, \ldots ,y_{t-1}) = {\rm {Generator}}([m_j, x^s_i]; \boldsymbol{\bar{h_r}})
\end{equation}

\section{Experiment}
\label{sec:exp}
In this section, we present our CogIEA dataset and some baselines for the three tasks with the results.

\subsection{Data Collection}
\label{sec:data}
\textbf{Basic Statistics} The dialogue dataset CogIEA is collected from DailyDialog \cite{DBLP:conf/ijcnlp/LiSSLCN17} and IEMOCAP \cite{DBLP:journals/lre/BussoBLKMKCLN08}, which contains 2,106 single-turn conversations. For the annotation, six emotion labels and seven intention labels are defined, as depicted in Sec. \ref{cogfr}. For example, for the utterances containing obvious request words, such as \textit{would like, ask for} and so on, annotators annotate them as request. For the utterances which express \textit{yes or no}, annotators annotate them as \textit{accept or reject}. Besides, another satisfaction label is added to indicate whether the action of the listener satisfied the intention of the speaker. Fig. \ref{fig:data} (c) shows the detailed statistics of the final dataset. We also count the distribution of emotion and intention to give a brief view of the dataset, as shown in Fig. \ref{fig:data} (a)(b).

\noindent
\textbf{Annotation Criteria} The annotators should first determine the intention of the speaker and the listener according to the dialogue. We have expanded the intention classification into seven categories based on \cite{DBLP:conf/ijcnlp/LiSSLCN17}, including \textit{request, suggest, command, accept, reject, question, inform}. As for the emotion, we have six emotional reaction in CogIEA (e.g., \textit{happy, content, neutral, sadness, anger, disgust}).
Then they are asked to identify whether the speaker's intention is satisfied by the listener's action. If the intention is satisfied, they will annotate positive emotion, otherwise, label as neutral or negative emotion. We assign each example to three annotators to annotate. When the results of two of them are consistent, the data is retained. Otherwise, an expert is invoked to make the final decision, ensuring the quality of data annotation. We removed the examples where no emotion or intention was selected. And each example is paid \$ 0.1.

\subsection{Experimental Setting}
We separate the CogIEA datasets into training (80\%), validation (10\%), test (10\%) sets. We train our models on the training sets, selecting the best performing model on validation set, for which we then evaluate test results. We set word embeddings to size of 300 and initialize them with Glove embeddings \cite{DBLP:conf/emnlp/PenningtonSM14}. In addition, we perform a grid search over the hyper-parameter settings (with a learning rate from \{0.01, 0.4\} for GRU or \{1e-5, 3e-5\} for PLMs, a batch size from \{16, 32\}, and epochs from \{3, 20\}). Models are trained to minimize the cross entropy with Adam \cite{DBLP:journals/corr/KingmaB14}. 
As for the evaluation metric, we use Precision (P), Recall (R) and F1 for task one and two. Automatic evaluation (BLEU, ROUGE) \cite{lin2004rouge,DBLP:conf/acl/PapineniRWZ02} and human evaluation \cite{DBLP:journals/corr/abs-2008-09075} are used in task three.

In the following, we provide some baselines and proposed modules \footnote{The data is available at: \small{~https://github.com/pengwei-iie/CogIntAc}} for three tasks, respectively.

\noindent
\textbf{GRU} \cite{DBLP:journals/corr/ChungGCB14}, where a GRU model is first used to construct the utterance representations, then with a MLP and softmax layer. The RNN is 2-layer GRU with 128 hidden units.

\noindent
\textbf{GRU+ATTENTION} \cite{Vaswani2017AttentionIA}, where a GRU model with the self-attention to model the semantic information of the utterance.

\noindent
\textbf{BERT} \cite{DBLP:conf/naacl/DevlinCLT19}, where the BERT is used as the contextual encoder, then with a MLP and softmax layer.

\noindent
\textbf{RoBERTa} \cite{DBLP:journals/corr/abs-1907-11692}, where it considers more data and bigger model for better performance. We introduce it for stronger baseline. 

\noindent
\textbf{BART} \cite{DBLP:conf/acl/LewisLGGMLSZ20}, a denoising autoencoder
for pretraining sequence-to-sequence models. We consider it and GPT-2 as the generator module.

\noindent
\textbf{GPT-2} \cite{Radford2019LanguageMA}, which is a 1.5B parameter Transformer for the generation task. The PLMs have the same hyperparameters given on the paper \cite{DBLP:conf/naacl/DevlinCLT19,DBLP:journals/corr/abs-1907-11692}.

\begin{figure*}[!]
	\centering
	\includegraphics[width=0.98\textwidth]{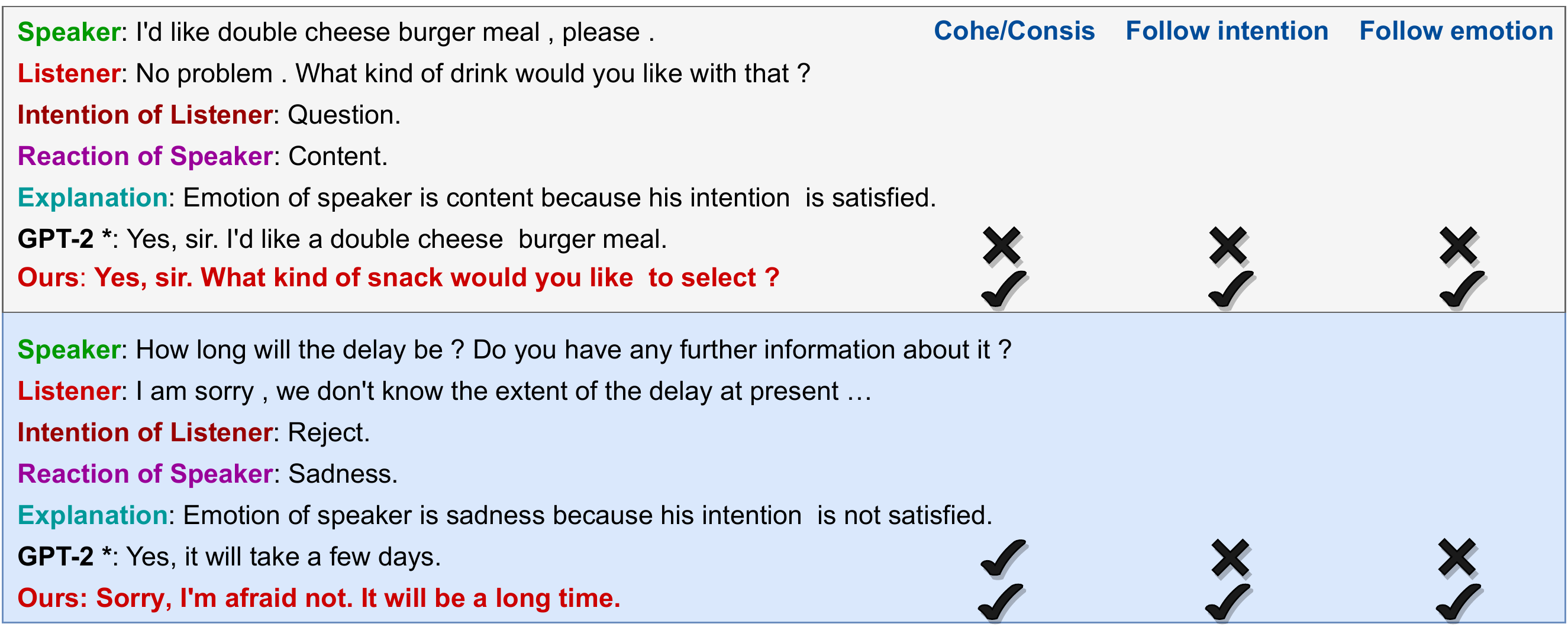}
	\caption{Two examples of generation approaches, we also give the explanation of the emotion prediction task.} 
	\label{fig:case}
\end{figure*}

\begin{table*}[h]
	\centering
	\resizebox{0.98\linewidth}{!}{
		\begin{tabular}{l|cccccc|cccc}
			\toprule
			& B-1 & B-2 & B-4 & R-1 & R-2 & R-L & Coherent & Consistent & Intention & Emotion \\
			\hline
			BART$_{large}$$^*$ \cite{DBLP:conf/acl/LewisLGGMLSZ20}& 23.55 & 2.61 & 0.16 & 14.47 & 2.55 & 13.30  & 2.61 & 2.44 & 2.07 & 1.96 \\
			GPT-2$_{large}$$^*$ \cite{Radford2019LanguageMA}  & 25.91 & 7.55 & 2.73 &  7.05 & 1.37 & 7.12  & 2.63 & 2.51 & 2.46 & 2.27 \\
			\hline
			BART$_{large}$ (ours) & \textbf{27.10} & 4.89 & 0.36 & \textbf{16.31} & \textbf{3.87} & \textbf{15.21}  & 2.67 & 2.53 & 2.33 & 2.30 \\
			GPT-2$_{large}$ (ours) & 25.91 & \textbf{8.67} & \textbf{3.08} &  8.87 & 2.07 & 8.66  & \textbf{2.71} & \textbf{2.76} & \textbf{2.68} & \textbf{2.53} \\
			
			\bottomrule
	\end{tabular}}
	\caption{Automatic evaluation and human evaluation of coherence, consistent (reported from 1-low to 3-high), intention, and emotion (reported from 1-low to 3-high). $^*$~ means with the only context as input. }
	\label{generation_ppl}
\end{table*}

\subsection{Experimental Results}
For thorough comparisons, we implement some baselines to the three tasks, including Action Abduction, Emotion Prediction, Action Generation.

\noindent
\textbf{Action Abduction of Speaker}
The Action Abduction Module, depicted in Fig. \ref{fig:model} (a), aims to obtain the intention of the speaker.
As shown in the first column of Table \ref{tab:main}, in all baselines, RoBERTa performs the best. After utilizing the IntDic, the performance has improved. It shows that the prior statistical knowledge is effective for the action abduction task to predict the intention.

\noindent
\textbf{Emotion Prediction of Speaker}
In this section, the results are shown in the second and third columns of Table \ref{tab:main}. In addition to IntDic, the fusion mechanism and multi-task learning also have an improvement, the model achieves the best results with all the components together. Another interesting point is that satisfaction prediction task scores highest because the model only needs to determine whether the action meets the intention, but when we increase the joint training for the tasks, we find that these tasks are also improved.

\noindent
\textbf{Action Generation of Listener}
The Action Generation Module generates the response of the listener. We ask testers to rate responses on the following criteria: (1) Coherence, or is the response on topic and strongly acknowledges the speaker, (2) Consistent, or does the response make logical sense given the context, (3) Follows intention, or does the response contain reasonable intention, and (4) Follows emotion, or does the response satisfy the speaker's emotional label. The results in Table \ref{generation_ppl} demonstrate that (1) compared with GPT-2$_{large}$$^*$ and BART$_{large}$$^*$, our model has achieved an improvement in both automatic and human evaluation. (2) the GPT-2 models achieve higher ratings for quality metrics of human evaluation. And the logic and coherence of the GPT-2 are relatively good, the expression form is relatively diversified.

\subsection{Qualitative Analysis}
We present dialogue cases and explanations of the emotion to demonstrate how our model performs on these tasks. 
As shown in Fig. \ref{fig:case}, for the first case, the speaker requests for \textit{burger meal} and intention of the listener is \textit{question}. The model only with context as input generates a sentence with a logical error \textit{I'd like a double ...} Our model produces a reasonable response which is consistent with the intention \textit{question}. Also, the response satisfies the speaker's emotional label. For the second case, the baseline outputs a reasonable utterance but neither the follows intention nor the follows emotion satisfies the label. Ours generates a better one. As for the explanation of the emotion, we give the generative template as demonstrated in Sec. \ref{model:emo}.

\section{Conclusion}
In this paper, we present the novel Cognitive framework of individual InterAction (CogIntAc) with the concept of \textit{a complete interaction} and \textit{interaction chain} from the perspective of psychology. Guided by the framework, we establish the interactive relationships by understanding and analyzing the relationships between elements in CogIntAc. Furthermore, we reconstruct a dataset CogIEA and introduce three tasks as well as the corresponding baseline models. 
Experimental results and qualitative analysis show that interactive action can be predictable by analyzing the nodes in the \textit{interaction chain}, and the analysis results can be explained reasonably. For the future work, we intend to expand into multi-turn of dialogues for further analysis.

\section*{Acknowledgment}
We thank all anonymous reviewers for their constructive comments and we have made some modifications. This work is supported by the National Natural Science Foundation of China (No.U21B2009).
%

\bibliographystyle{IEEEtran}
\bibliography{mybibfile}


\end{document}